\documentclass[showpacs,aps,preprint,superscriptaddress,floatfix,nofootinbib]{revtex4}

\usepackage{epsf} \usepackage{ulem} \usepackage{graphicx}
\usepackage[usenames]{color}
 \def\bsg{B\rightarrow X_s
  \gamma} \def\bpkz{B^0\rightarrow \phi K^0} \def\bpk{B\rightarrow
  \phi K} \def\bpks{B^0\rightarrow \phi K_S} \def\bpkp{B^+\rightarrow
  \phi K^+} \def\bpkm{B^-\rightarrow \phi K^-} 
\def\beq{\begin{equation}} \def\eeq{\end{equation}}
\def\beqa{\begin{eqnarray}} \def\eeqa{\end{eqnarray}}
\def\ba{\begin{array}} \def\ea{\end{array}}

\begin{document}

\title{$B^0\rightarrow\phi K_S$ in SUGRA models with CP violations}

\author{R. Arnowitt} \email{arnowitt@physics.tamu.edu}
\affiliation{Center For Theoretical Physics, Department of Physics,
  \\Texas A$\&$M University, College Station, TX, 77843-4242, USA}

\author{B. Dutta} \email{duttabh@yogi.phys.uregina.ca}
\affiliation{Department of Physics, University of Regina, \\Regina SK,
  S4S 0A2, Canada}

\author{B. Hu} \email{b-hu@physics.tamu.edu} \affiliation{Center For
  Theoretical Physics, Department of Physics, \\Texas A$\&$M
  University, College Station, TX, 77843-4242, USA}

\date{\today}

\begin{abstract}
We examine the $\bpk$ decays within the framework of SUGRA models 
making use of the improved QCD factorization method of Beneke et al. which 
allows calculations of non-factorizable contributions. All other experimental 
constraints ($\bsg$, neutron and electron electric dipole moments, dark 
matter constraints, etc.) are imposed. We calculate the CP violating parameters 
$S_{\phi K_S}$, $C_{\phi K_S}$ and $\mathcal{A}_{\phi K^{\mp}}$ as well as the branching ratios (BR) of $B^0$ and 
$B^\pm$, $\mbox{Br}[\bpk]$. We find for the Standard Model(SM) and mSUGRA it is not 
possible to account for the observed $2.7\sigma$ deviation between $S_{\phi K_S}$ and 
$S_{J/\Psi K_S}$. In general the BRs are also in $3\sigma$ disagreement with 
experiment, except in the parameter region where the weak annihilation terms 
dominate the decay (and hence where the theory is least reliable). Thus if 
future data confirm the current numbers, this would represent the first 
significant breakdown of both the SM and mSUGRA. We show then that adding a 
SUGRA non-universal A soft breaking left-right term mixing the second and 
third generations in either the down or up quark sector, all data can be 
accommodated for a wide range of parameters. The full 6x6 quark mass matrices 
are used and the SUSY contributions calculated without approximation.      

\end{abstract}


\maketitle

\section{Introduction}

Rare decay modes of the $B$ meson are important places to test the
Standard Model (SM) and to look for new physics. However, large
theoretical uncertainties in the calculations of exclusive
non-leptonic $B$ decays make it difficult to extract useful
information from experimental data. Nevertheless, CP asymmetries of
neutral $B$ meson decays into final CP eigenstates, i.e.,
$B\rightarrow\phi K_S$ and $B\rightarrow J/\Psi K_S$, are uniquely
clean in their theoretical interpretations. Among these decay modes,
$\bpks$ is induced only at the one loop level in the SM and hence is a
very promising mode to see the effects of new physics. In the SM, it
is predicted that the CP asymmetries of $\bpks$ and $B\rightarrow
J/\Psi K_S$ should measure the same $\sin 2\beta$ with negligible
$O(\lambda^2)$ difference \cite{worah}.  On the other hand, the BaBar
and Belle measurements \cite{babar02a, belle02, babar02b, babar03}
show a $2.7\sigma$ disagreement between $S_{\phi K_S}$ and $S_{J/\Psi
  K_S}$ \cite{babar03}\footnote{After submitting this work new data from Belle
\cite{BelleNew} gave a value of $S_{\phi K_S}  = -0.96 \pm 0.5^{+0.09}_{-0.11}$ (a 
$3.5\sigma$ deviation from the Standard Model) and preliminary analysis of new data 
from BaBar \cite{Browder} gave $S_{\phi K_S}  = +0.45 \pm 0.43\pm 0.07$. Belle and 
BaBar would then disagree by $2.1\sigma$ and if one averages the new values one 
obtains  \cite{Browder} $S_{\phi K_S}  = -0.15\pm 0.33$ which is again $2.7\sigma$ 
from the Standard Model.}: 
\beqa\label{eq01}
S_{J/\Psi K_S} & = & 0.734 \pm 0.055,\nonumber\\
S_{\phi K_S} & = & -0.38 \pm 0.41.  \eeqa while $S_{J/\Psi K_S}=\sin
2\beta_{J/\Psi K_S}$ (which is a tree level process) is in excellent
agreement with Buras' SM evaluation, $\sin
2\beta=0.715^{+0.055}_{-0.045}$ from the CKM matrix \cite{Buras}. In
addition, the branching ratios (BRs) and the direct CP asymmetries of
both the charged and neutral modes of $\bpk$ have also been measured
\cite{babar02a, belle02, babar02b, babar03}\footnote{Our average of $\mbox{Br}[\bpkp]$ only includes BaBar and Belle since CLEO \cite{footnote} is
$2.3\sigma$ away. $\mbox{Br}[\bpkp]$ would become $(9.4 \pm 0.9) \times 10^{-6}$
if CLEO is included.}:
\beqa\label{eq01a} \mbox{Br}[\bpks] & = & (8.0\pm 1.3) \times
10^{-6},\nonumber\\\mbox{Br}[\bpkp] & = & (10.9 \pm
1.0) \times 10^{-6},\\\label{eq01b} C_{\phi K_S} & = & -0.19 \pm 0.30,\nonumber\\
\mathcal{A}_{CP}(\bpkp) & = & (3.9 \pm 8.8 \pm 1.1) \% .  \eeqa
In general, any model should explain all these data. In particular,
the relatively small uncertainties in the BRs of $\bpkp$ and $\bpks$
need to be considered in the analysis since they are highly
correlated and both are based on the $b\rightarrow s$ transition.
In the SM, $\mathcal{A}_{CP}(\bpkp)$ is small and agrees with
(\ref{eq01b}). So this direct CP asymmetry result plays an important
role in constraining the new physics contribution which might explain
the discrepancy between $S_{\phi K_S}$ and $S_{J/\Psi K_S}$.

The discrepancy between $S_{\phi K_S}$ and $S_{J/\Psi K_S}$ has been
discussed in some recent works \cite{kagan, Hiller, ADatta, Ciuchini02a, Dutta02a,
  Khalil, Baek, Chiang, Kundu, Agashe, Kane03, Chak, csh} in the framework of
SUSY models, especially in the minimal supersymmetric standard model
(MSSM) with the mass insertion method \cite{mia}. Although these works can
provide useful constraints on certain off-diagonal terms of squark
mass matrices at low energy, we find that it is very interesting to
investigate this problem in the context of grand unified theory (GUT)
models. Since GUT models explain a number of phenomena at low energy
by a few well motivated parameters at the GUT scale, various
experimental measurements get correlated in this framework. Among
supersymmetric GUT models, the R-parity conserved SUGRA model is one
of the most favored models since it provides a natural explanation of
the dark matter problem. The minimal SUGRA model (mSUGRA) with
R-parity conservation has been investigated extensively because of its
predictive power that comes from the fact that it depends on only a
few new parameters. Unlike the MSSM, which is hard to be constrained
due to its more than 100 new parameters (including 43 CP violating phases),
the parameter space of the minimum SUGRA model has 4 parameters and 4
phases. This parameter space has several experimental constraints,
i.e. $b\rightarrow s + \gamma$, neutron and electron electric dipole
moments (EDM), LEP bounds and relic density
measurements\cite{Arnowitt02a, cnst} etc.. In this paper, we
examine the observed discrepancy between $S_{\phi K_S}$ and
$S_{J/\Psi K_S}$ in the context of SUGRA models including mSUGRA and
models with non-universalities. We consider all relevant experimental
constraints in our calculation.

\begin{figure}[t]
  \scalebox{0.80}[0.80]{\includegraphics{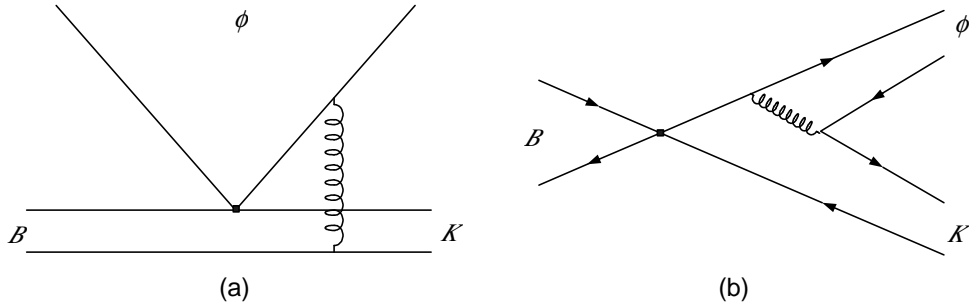}}
        \caption{\label{Feyn}Hard spectator scattering diagram (a) and weak annihilation diagram (b). In (a) the gluon can connect the spectator with either $\phi$ quark and in (b) the gluon can originate from any $B$ quark or $K$ quark.}
\end{figure}

The calculation of $\bpk$ decays involve the evaluation of the matrix
elements of related operators in the effective Hamiltonian, which is
the most difficult part in this calculation. However, the newly
developed QCD improved factorization (BBNS approach) \cite{bbns}
provides a systematic way to calculate the matrix elements of a large
class of $B$ decays with significant improvements over the old
factorization approach (naive factorization). It allows a QCD
calculation of ``non-factorizable'' contributions and model
independent predictions for strong phases which are important in the
theoretical evaluation of the direct CP asymmetries of $B$ decays,
e.g. for $\bpkm$, whose experimental result is given in (\ref{eq01b}).
We adopt the QCD improved factorization in our $\bpk$ calculations.
Recently Du et al. \cite{Du02a, Du02b,zhu} have published an improved
calculation of $B\rightarrow PV$ decays. We followed here their
calculational techniques \cite{Du02a} which are based on the original
work \cite{bbns} of Beneke, et al. While the BBNS approach is an
important advance in calculating $B$ decays, it is not completely
model independent.  In the BBNS approach the hard gluon (H) and annihilation (A)
diagrams (see Fig.\ref{Feyn}) contain infrared divergences which are
parameterized by an amplitude $\rho_{H,A}$ (with $\rho_{H,A}\leq 1$)
and a phase $\phi_{H,A}$. (More details can be found in \cite{bbns} and
\cite{Du02a}.) If the effects of these terms are small, the
theoretical predictions are well defined. However, if these terms are
large or dominant, the theory becomes suspect. We will see below that
$S_{\phi K_s}$ is essentially independent of the infrared divergent
terms, though the branching ratios can become sensitive to $\rho_A$
and $\phi_A$.

In this work, we first examine the mSUGRA model which is universal at
the GUT scale and then consider  non-universal terms. Although 
non-universalities may cause serious problems in some flavor changing
processes, e.g. $K^0 - \bar{K^0}$ mixing and $b\rightarrow s +
\gamma$, we will show that the off-diagonal terms in the A parameter
soft-breaking terms can satisfy all experimental data including the
$2.7\sigma$ deviation between $S_{\phi K_S}$ and $S_{J/\Psi K_S}$.  We
calculate the BR of $\bpkm$ which is highly correlated with $\bpkz$
and calculate the CP asymmetry of this mode in the allowed parameter
space. We also calculate the CP asymmetry of the $b\rightarrow
s\gamma$ decay mode.

This paper is organized as following: In Sec.~2, we give a brief
description of the CP asymmetry of $\bpks$ and the QCD factorization
technique used in this paper. Then we discuss the SUGRA model 
and its contributions to $\bpks$ in Sec.~3. Some detailed discussion on experimental
constraints implemented in our analysis are given in Sec.~4 and the 
Standard Model predictions are discussed in Sec.~5. Sec.~6 is devoted
to our results for the mSUGRA model, after which we proceed to the
models with non-universalities in Sec.~7 and we give conclusions in
the last section.

\section{CP asymmetry of $\bpk$ decays}

The time dependent CP asymmetry of $B\rightarrow \phi K_S$ is
described by:
\begin{eqnarray}\label{eq02}
\mathcal{A}_{\Phi K_S}(t) & \equiv &
\frac{\Gamma(\overline{B}^0_{\mbox{phys}}(t) \rightarrow \phi K_S) -
\Gamma(B^0_{\mbox{phys}}(t) \rightarrow \phi
K_S)}{\Gamma(\overline{B}^0_{\mbox{phys}}(t) \rightarrow \phi K_S) +
\Gamma(B^0_{\mbox{phys}}(t) \rightarrow \phi K_S)}\nonumber\\
& = & -C_{\phi K_S}\cos (\Delta m_B t) + S_{\phi K_S}\sin (\Delta m_B t)
\end{eqnarray}
where $S_{\phi K_S}$ and $C_{\phi K_S}$ are given by \beq\label{eq03}
S_{\phi K_S}=\frac{2\,\mbox{Im}\lambda_{\phi K_S}}{1+|\lambda_{\phi
    K_S}|^2}\;,\;\;\;\;C_{\phi K_S}=\frac{1-|\lambda_{\phi
    K_S}|^2}{1+|\lambda_{\phi K_S}|^2}, \eeq and $\lambda_{\phi K_S}$
can be written in terms of decay amplitudes: \beq\label{eq04}
\lambda_{\phi K_S} = - e^{-2i\beta}
\frac{\overline{\mathcal{A}}(\overline{B}^0\rightarrow \phi
  K_S)}{\mathcal{A}(B^0\rightarrow \phi K_S)} \eeq In our analysis, we
find that the SUSY contributions to $B_d - \overline{B}_d$ mixing are
small, and so from now on we will use the standard definition for
$\beta$: \beq\label{eq05} \beta \equiv \mbox{arg} \left( \frac{V_{cd}
    V_{cb}^\star}{V_{td} V_{tb}^\star} \right).  \eeq Within the SM,
$\sin 2\beta$ can be measured by $S_{J/\Psi K_S}$. The current
experimental result is given in Eq. (\ref{eq01}). Since
$B^0\rightarrow J/\Psi K_S$ decay is dominated by the SM tree level
contribution, we expect that in our analysis the new physics will not
affect the SM prediction for $\sin 2\beta$ from $B\rightarrow J/\Psi
K_S$. As a consequence, we further assume that the current SM fit for
the CKM matrix will not be affected by models discussed in this paper.

The CP asymmetry of charged $\bpk$ decay is defined as
\beq\label{eq05a} \mathcal{A}_{\phi K^{\mp}} \equiv \frac{\Gamma(B^-
  \rightarrow \phi K^-) - \Gamma(B^+ \rightarrow \phi K^+)}{\Gamma(B^-
  \rightarrow \phi K^-) + \Gamma(B^+ \rightarrow \phi K^+)} =
\frac{|\lambda_{\phi K^{\mp}}|^2-1}{|\lambda_{\phi K^{\mp}}|^2+1} \eeq
where \beq\label{eq05b} \lambda_{\phi K^{\mp}} =
\frac{\overline{A}(B^- \rightarrow \phi K^-)}{A(B^+ \rightarrow \phi
  K^+)}. \eeq

From the above discussion, it is clear that our theoretical
predictions for the experimental observables, e.g.  $S_{\phi K_S}$,
$C_{\phi K_S}$ and $\mathcal{A}_{\phi K^{\mp}}$, depend on the
evaluation of decay amplitudes where the effective Hamiltonian plays
an important role. The Effective Hamiltonian for $\bpk$ in the SM is
\cite{bbns}: \beq\label{eq06} \mathcal{H}_{eff} = \frac{G_F}{\sqrt{2}}
\sum_{q=u,\,c} V_{qb}V_{qs}^\star \left[ C_1O_1^q + C_2O_2^q +
  \sum_{k=3}^{10} C_k(\mu)O_k(\mu) + C_{7\gamma}O_{7\gamma} +
  C_{8g}O_{8g} \right] + \mbox{h.c.}  \eeq where the Wilson
coefficients $C_i(\mu)$ can be obtained by running the RGE from the
weak scale down to scale $\mu$. The definitions of the operator
$O_i$'s in the SM can be found in \cite{bbns}. The SUSY contributions
will bring in new operators $\tilde{O}_i$'s which can be obtained by
changing $L\leftrightarrow R$ in the SM operators. We use $\tilde{C}_i$ to denote the Wilson coefficient of $\tilde{O}_i$. The decay amplitude
of $B\rightarrow M_1M_2$ can be expressed in terms of the matrix
elements of $O_i$'s, $\langle M_1M_2|O_i|B \rangle$. We evaluate these matrix
elements in the QCD improved factorization technique. The necessary
expressions can be found in \cite{bbns,Du02a}.

Using the above Hamiltonian the amplitude of $\bpk$ is:
\beq\label{eq06a} \mathcal{A}(B\rightarrow \phi K) =
\mathcal{A}^f(B\rightarrow \phi K) + \mathcal{A}^a(B\rightarrow \phi
K) \eeq where $\mathcal{A}^f$ is factorized amplitudes which can be
written as \cite{Du02a} \beq\label{eq06b} \mathcal{A}^f(B\rightarrow
\phi K) = \frac{G_F}{\sqrt{2}} \sum_{p=u,\,c} \sum_{i}
V_{pb}V_{ps}^\star a^p_i \langle \phi K|O_i|B \rangle_f, \eeq and $\mathcal{A}^a$ is
the weak annihilation decay amplitudes \cite{Du02a}: \beq\label{eq06c}
\mathcal{A}^a(B\rightarrow \phi K) = \frac{G_F}{\sqrt{2}} f_B f_\phi
f_K \sum V_{pb}V_{ps}^\star b_i. \eeq 
The matrix elements $\langle \phi K|O_i|B \rangle_f$ in
Eq.(\ref{eq06b}) are the factorized hadronic matrix elements
\cite{Ali}.  $a_i$'s and $b_i$'s contain the Wilson coefficients.
Explicit expressions for them, as well as for
$\mathcal{A}^a(B\rightarrow \phi K)$, can be found in \cite{bbns} and
\cite{Du02a}.

In our discussion, the dominant SUSY contributions occur through
$O_{7\gamma}$ and $O_{8g}$ and we calculate these new SUSY
contributions in the SUGRA framework.




\section{SUGRA models}

The SUGRA model at the GUT scale can be described by its
superpotential and soft-breaking terms: \beqa\label{eq07}
W & = & Y^U Q H_2 U + Y^D Q H_1 D + Y^L L H_1 E + \mu H_1 H_2 \nonumber\\
\mathcal{L}_{\mbox{\small soft}} & = & - \sum_i m_i^2 |\phi_i|^2 - \left({1 \over 2} \sum_{\alpha} m_{\alpha} \bar{\lambda}_\alpha \lambda_\alpha + B \mu H_1 H_2 \right. + \nonumber\\
& & \left. \left(A^U Q H_2 U + A^D Q H_1 D + A^L L H_1 E \right) +
  \text{h.c.} \right).  \eeqa Here $Q$, $L$ are the left handed quark
and lepton doublets, $U$, $D$ and $E$ are the right handed up, down
and lepton singlets and $H_{1,2}$ are the Higgs doublets. In the
minimal picture, the mSUGRA model contains a universal scalar mass
$m_0$, a universal gaugino mass $m_{1/2}$ and the universal cubic
scalar A terms: \beq\label{eq08}
m_i^2=m_0^2,\;\;\;m_{\alpha}=m_{1/2},\;\;\;A^{U,D,L}=A_0Y^{U,D,L}.
\eeq This model contains four free parameters and a sign: $m_0$,
$m_{1/2}$, $A_0$, $\tan\beta=\langle H_2 \rangle/\langle H_1 \rangle$ and the sign of $\mu$.

However, the parameters $m_{1/2}$, $\mu$ and $A$ can be complex and
their phases can be $O(1)$. In order to accommodate the experimental bounds 
on the electron and neutron electric dipole moments (EDMs) without fine tuning 
phases we extend mSUGRA by allowing the gaugino masses at $M_G$ to have arbitrary 
phases. This model has been extensively studied in the literature 
\cite{edm1,edm2,AAB1}. Thus the SUSY parameters with phases at the GUT scale 
are $m_{i}=|m_{1/2}|e^{i\phi_i}$ i=1,2,3 (the gaugino masses are for the $U(1)$,
$SU(2)$ and $SU(3)$ groups), $A_0=|A_0|e^{i\alpha_A}$ and 
$\mu=|\mu|e^{i\phi_\mu}$. However, we can set one of the 
gaugino phases to zero and we choose $\phi_2=0$. Therefore, we are left with four 
phases. The EDMs of the electron and neutron can now allow the 
existence of large phases in the theory \cite{edm1,edm2,AAB1}. In our calculation, 
we use $O(1)$ phases but calculate the EDMs to make sure that current bounds are 
satisfied.

We evolve the above parameters from the GUT scale down to the weak
scale using full matrix RGEs. Since the $b\rightarrow s$ transition
is a generation mixing process, it is necessary to use the full
$6\times 6$ matrix form of squark mass matrices in the calculation. We
perform the calculation of SUSY contributions without any
approximation.

We also include the one loop correction to bottom quark mass from
SUSY \cite{Car}, which is important in the calculation of SUSY
contributions to the Wilson coefficients of the operator $O_{7\gamma}$
and $O_{8g}$ and consequently affects the calculations of
$B\rightarrow X_s \gamma$ and $\bpk$ decays. We now discuss the
experimental constraints in the next section.

\section{Parameter space and Experimental Bounds}
In this section we review all the experimental constraints considered
in our analysis and briefly discuss their effects and importance.

\subsection{$B\rightarrow X_s \gamma$}
We use a relatively broad range for the branching ratio of
$B\rightarrow X_s \gamma$ \cite{Alam} to take into account the
uncertainty in the theoretical calculation of $\bsg$ ($\pm 0.3 \times 10^{-4}$):
\begin{equation}\label{eq09}
2.2  \times 10^{-4}< \mbox{Br}(B\rightarrow X_s \gamma) < 4.5 \times 10^{-4}.
\end{equation}
The SM prediction for the $\mbox{Br}[\bsg]$ is very close to the
measured value \cite{bsg}, so the $b\rightarrow s$ transition in any
new physics is strongly constrained. Since the $\bpk$ decay also
depends on the $b\rightarrow s$ transition, the $\mbox{Br}[\bsg]$
constraint needs to be implemented in any analysis of the $\bpk$
decay. Besides the BR of $\bsg$, we also consider in this work the direct CP asymmetry of $\bsg$ for the existing of CP violating phases. The experimental measurement from CLEO gives \cite{bsgcp}:
\begin{equation}\label{eq09a}
\mathcal{A}_{b\rightarrow s + \gamma} = (-0.079 \pm 0.108 \pm 0.022)(1.0 \pm 0.030)
\end{equation} or at 90\% confidence level, 
$-0.27 <\mathcal{A}_{b\rightarrow s + \gamma} < +0.10$.

\subsection{The relic density}
The recent WMAP result gives \cite{wmap}
\begin{equation}\label{eq10}
        \Omega_{CDM} h^2 = 0.1126^{+0.008}_{-0.009}.
\end{equation}
We implement this bound at the $2\sigma$ level in our
calculation:$0.094<\Omega h^2<0.129$. We also notice that when
non-universal terms are present, new annihilation channels may arise
and they are different from the usual mSUGRA $\tilde{\tau}-\tilde{\chi}^0$
co-annihilation channel.

\subsection{$K^0-\overline{K^0}$ mixing}
It has been shown that $K^0-\overline{K^0}$ mixing can significantly
constrain certain flavor changing sources in SUSY models \cite{Ciu}.
The current experimental bound for $\Delta M_K$ is \cite{pdg}:
\begin{equation}\label{eq11}
\Delta M_K = 3.490 \pm 0.006 \times 10^{-12}\;\mbox{MeV}
\end{equation}
In non-universal SUGRA models, even if non-universal terms between the
first two generations are not present at the GUT scale, $\Delta M_K$
can become large. This happens because the degeneracy between the
first two generations can get broken by other non-universal terms via
the RGEs. For example, the $A_{32}$ terms in the trilinear
coupling matrices $A$ in Eq.(\ref{eq07}) can give rise to new
contribution to the $m^2_{22}$ term via $\delta m^2_{22} \propto {1 \over
  16 \pi^2} A_{23}A_{32}Log[M_{GUT}/M_{weak}]$. Therefore, it is
important to pay attention to $\Delta M_K$ even when there is no apparent direct
source producing large contributions to $\Delta M_K$.
 
\subsection{Neutron and electron electric dipole moments}
Neutron and electron electric dipole moments (EDM) can arise in any
model with new CP violating phases.  In SUSY models, an electron EDM
arises from diagrams involving intermediate chargino-sneutrino states
and intermediate neutralino-selectron states (for more details, see \cite{edm1,edm2,AAB1}).
The current experimental bounds on 
neutron and electron EDMs are \cite{pdg}:
\begin{equation}\label{eq12}
                d_{n} < 6.3 \times 10^{-26} e\,cm,\;\;\;d_{e} < 0.21 \times 10^{-26} e\,cm
\end{equation}

There are other important phenomenological constraints considered,
e.g.  bounds on masses of SUSY particles and the lightest Higgs
($m_h\geq 114$ GeV).

\section{$\bpk$ decays in the Standard Model}
We first discuss $\bpk$ decays in the SM. The largest theoretical
uncertainties in this calculation come from weak annihilation
diagrams which mostly depend on the divergent end-point integrals
$X_A$ parameterized in the form \cite{bbns,Du02a} \beq\label{eq13} X_A
= (1+\rho_A e^{i\phi_A})\ln
\frac{m_B}{\Lambda_h},\,\Lambda_h=\Lambda_{QCD},\,\rho_A\leq 1. \eeq
Hard spectator processes contain similar integrals $X_H$ which are
parameterized in the same way. However, uncertainties
from the hard spectator calculation are much smaller than those from
the weak annihilation for this decay, so we will mainly concentrate on
the later. These weak annihilation contributions depend also on the
strange quark mass, $m_s$, through the chirally enhanced factor
$\kappa_\chi$: \beq\label{eq13a} \kappa_\chi =
\frac{2m_K^2}{m_b(m_s+m_q)} \eeq where $m_q$ is  $m_d$ or $m_u$.

\begin{figure}[H]
  \scalebox{0.85}[0.85]{\includegraphics{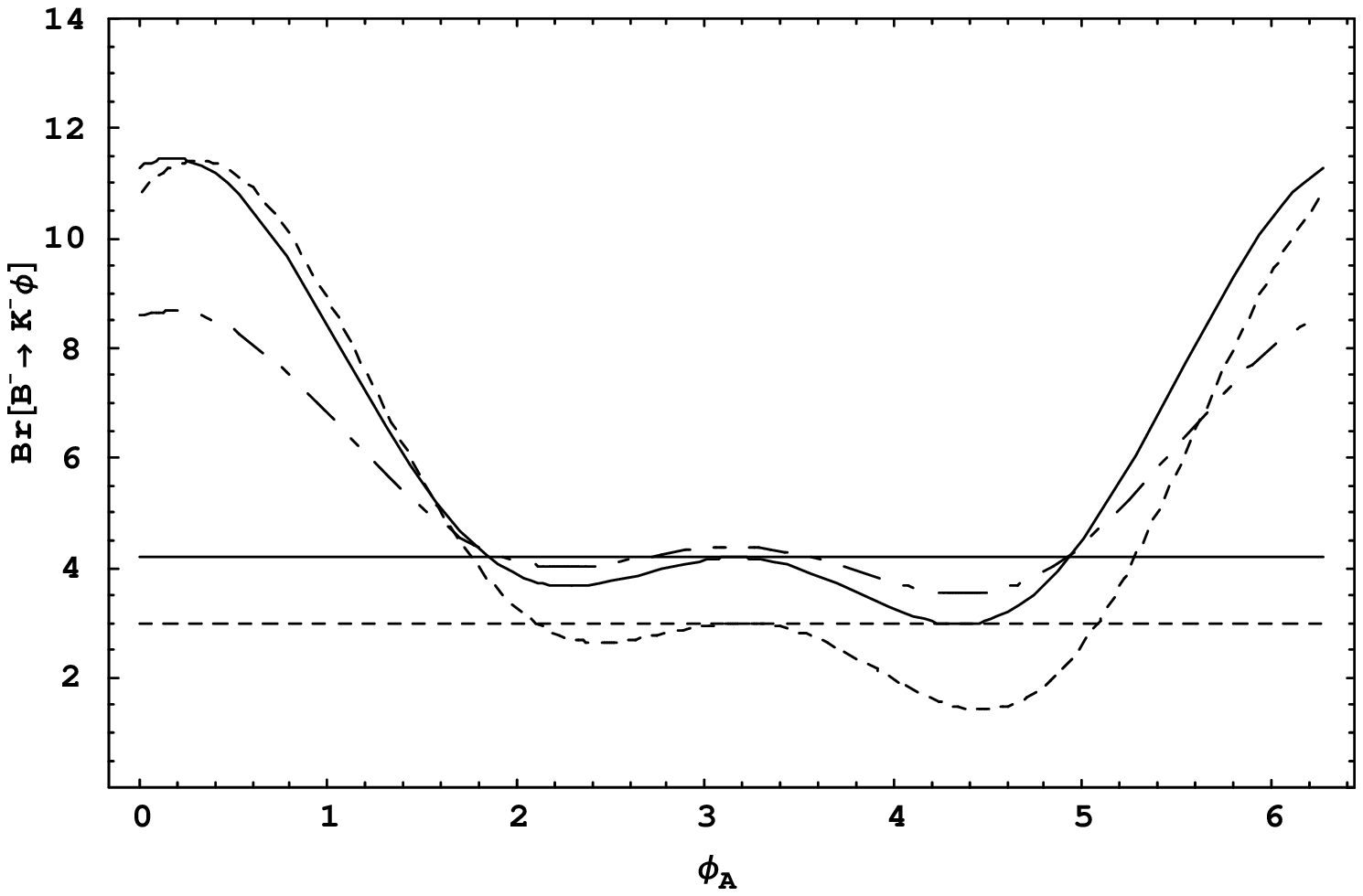}}
        \caption{\label{phi}Branching ratio of $\bpkm$  at $\rho_A = 1$. The solid curve corresponds to $\mu = m_b$, dashed curve for $\mu = 2.5\,\mbox{GeV}$ with $m_s(2\,\mbox{GeV}) = 96\, \mbox{MeV}$ and the dot-dashed curve for $\mu = m_b$ with $m_s(2\, \mbox{GeV}) = 150\,\mbox{MeV}$. The two  straight lines correspond to the cases without weak annihilation.}
\end{figure}
\begin{figure}[H]
  \scalebox{0.80}[0.80]{\includegraphics{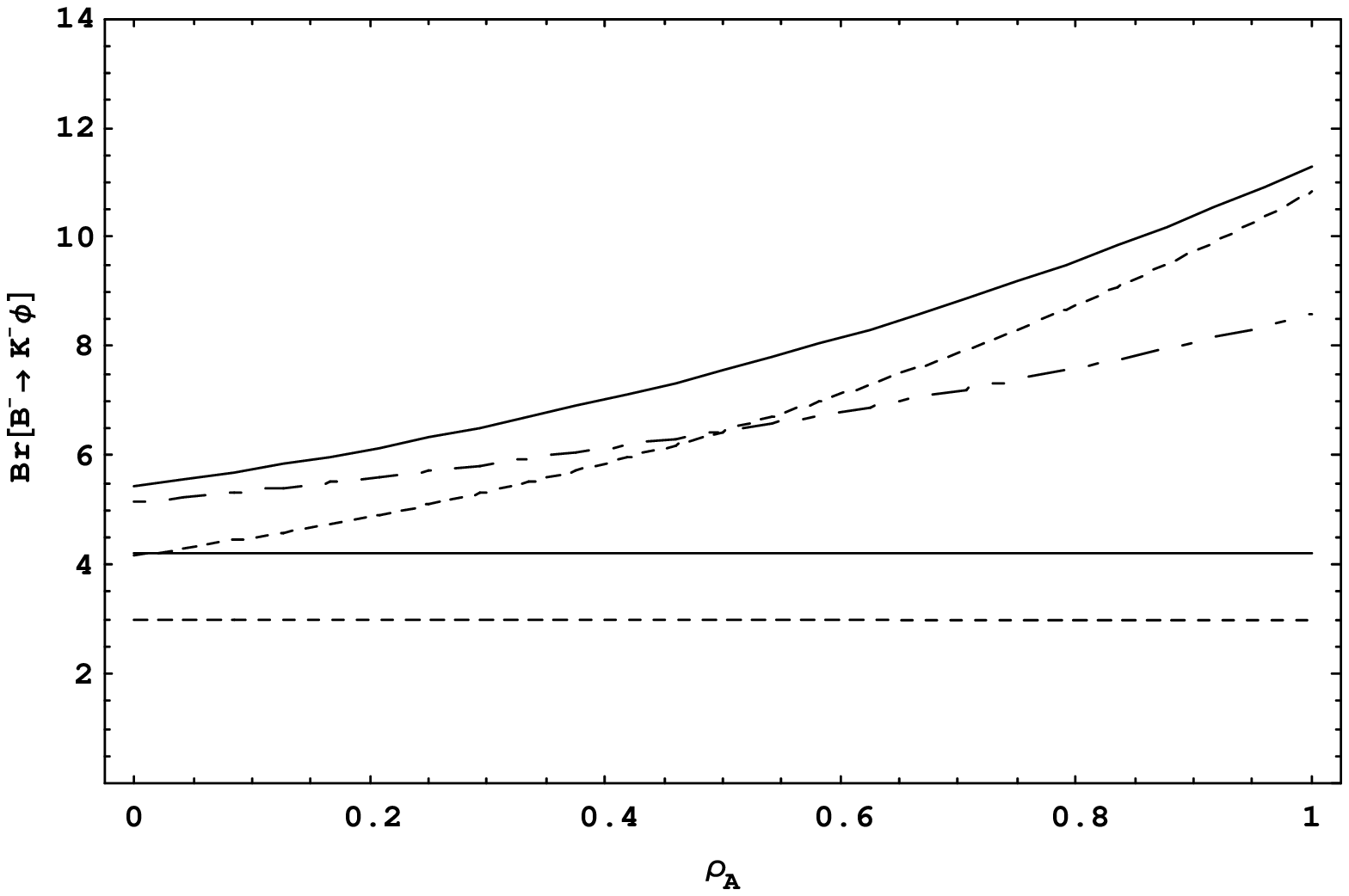}}
        \caption{\label{rho}Branching ratio of $\bpkm$  at $\phi_A = 0$. The solid curve corresponds to $\mu = m_b$, dashed curve for $\mu = 2.5\,\mbox{GeV}$ with $m_s(2\,\mbox{GeV}) = 96\, \mbox{MeV}$ and the dot-dashed curve for $\mu = m_b$ with $m_s(2\, \mbox{GeV}) = 150\,\mbox{MeV}$. The two  straight lines correspond to the cases without weak annihilation.}
\end{figure}

In Fig.\ref{phi} we show the dependence of the branching ratio of
$\bpkm$ mode on $\phi_A$ and $m_s$ for $\rho_A = 1$. Fig.\ref{rho}
shows the dependence of the BR on $\rho_A$.  Similar graphs can be
obtained for $\bpkz$.  Since $\mathcal{A}_{\phi K^{\mp}}$ in the SM
(Eq.(\ref{eq05a})) is small, we can compare $\mbox{Br}[\bpkm]$ with
the experimental measurement of $\mbox{Br}[\bpkp]$ given in
Eq.(\ref{eq01a}). Before we discuss the graphs, we first list our
parameters: $\rho_H = 1$, $\phi_H = -68^{\circ}$, $f_B =
180\,\mbox{MeV}$, $f_\phi = 233\,\mbox{MeV}$, $f_K = 160\,\mbox{MeV}$
and $F^{BK} = 0.34$.  The CKM matrix elements can be obtained through the Wolfenstein
parameterization \cite{Wp} with $A=0.819$, $\lambda = 0.2237$,
$\rho=0.224$ and $\eta=0.324$. The integral $\int_0^1 (\Phi_B (\xi) /\xi) d\xi
= m_B / \lambda_B$ where $\Phi_B$ is the B meson light-cone distribution amplitude is parameterized by $\lambda_B=(0.35\pm0.15)$ GeV.  
For $\mu=2.5\,\mbox{GeV}$ we use $\lambda_B=0.2\,\mbox{GeV}$, and for $\mu=m_b$  we
use $\lambda_B=0.47\,\mbox{GeV}$. In addition, we always use asymptotic
forms of the meson light-cone distribution amplitudes \cite{bbns,
  Du02a}.  If not mentioned, we will use the above parameters in later
calculations.

In both figures, we give results for two different scales and two
different $m_s$ values, i.e., $\mu=m_b$ by solid lines
($m_s(2\,\mbox{GeV}) = 96\, \mbox{MeV}$) and the dot-dashed lines
($m_s(2\,\mbox{GeV}) = 150\, \mbox{MeV}$) and $\mu=2.5\,\mbox{GeV}$ by dashed
lines ($m_s(2\,\mbox{GeV}) = 96\, \mbox{MeV}$). One can see that the
scale dependence is not significant. The straight lines correspond to
the branching ratios neglecting the weak annihilation contribution.
Comparing Figs \ref{phi} and \ref{rho}, we see that a large branching
ratio comparable to the experimental value is obtained in the region
$\rho_A \simeq 1$ and $\phi_A\simeq 0$(or 2$\pi$). However, in this
region the weak annihilation diagrams dominate the branching ratio and
thus the theory is most suspect. In the remaining part of the
parameter space, where the weak annihilation effects are small and the
theory is presumably reliable, the SM prediction of the branching
ratio is small, about $3\sigma$ below the experimental value. We
conclude therefore that where the theory is reliable the SM is in
significant disagreement with the experimental value of the
$\mbox{Br}[B^+\rightarrow\phi K^+]$, and in order to obtain a SM value
in accord with the experiment one must use parameters where the theory
is least reliable. A similar result holds for the
$\mbox{Br}[B^0\rightarrow\phi K_s]$. Here theory predicts a branching
ratio about 10\% smaller than for $B^+\rightarrow\phi K^+$ (in accord
with the experimental values of Eq.(\ref{eq01a})) but the SM can
achieve this only in the region where the weak annihilation processes
dominate.

The dot-dashed line, in Fig.\ref{phi}, corresponds to a larger value
of $m_s$ and we see that the BR is very sensitive to $m_s$ only in the
large annihilation region. The region with sufficiently large
annihilation to accommodate the data decreases as $m_s$ increases,
since the annihilation amplitude then decreases, as can be seen from
Eq.(\ref{eq13a}).


\section{mSUGRA}
Before we proceed to present our results, let us first mention the
values of the parameters used in our calculation of $\bpk$ decay
amplitudes in SUGRA models since the BRs depend sensitively on them.
We use $\rho_{A,H} =1$, $\phi_{A,H} = -68^{\circ}$,
$m_s(2\,\mbox{GeV}) = 122.5\, \mbox{MeV}$ and a CKM fit giving $\sin 2
\beta = 0.73$ and $\gamma = 59^{\circ}$. (If we increase $\gamma$, the
BR decreases, e.g. for $\gamma = 79^{\circ}$, the SM BR decreases by
$\sim 2$\%.) The SM BR is $4.72 \times 10^{-6}$ and the weak
annihilation contribution is small ($\sim 10$ \%).



In Table \ref{ms1}, we give the numerical results for two different
values of the mSUGRA parameter $\tan\beta$ cases i.e. $\tan\beta=40$
and 10 and for different $m_{1/2}$ and $A_0$ in the small weak
annihilation case. For simplicity, we set $\alpha_A = \pi$ in the
calculation of Table I. We use large phases for other parameters but
still satisfy the EDM constrains. For example, for $m_{1/2}=400$ GeV
and $A_0 = 800$ GeV with $\tan\beta = 40$, we find that $\phi_1 =
70^{\circ}$, $\phi_3 = 33^{\circ}$ and $\phi_\mu = -13^{\circ}$ (at
the weak scale) satisfy the EDM constraints (for reasons discussed in
detail in \cite{AAB1}). The phase $\alpha_A$ has a very small effect
on $S_{\phi K_S}$ and this effect becomes smaller as the magnitude of
$A$ decreases. Thus a different value of $\alpha_A$, in the above fit,
can change $S_{\phi K_S}$ by $\pm 4\%$ for $A_0 = 800$ GeV. This
change is even smaller for smaller $A_0$, e.g. for $A_0 = 200$ GeV,
the change in $S_{\phi K_S}$ is less than $2$\%. So far we have not
specified $m_0$ for our results. The values of $m_0$, for different
$m_{1/2}$ and $A_0$, are chosen such that the relic density constraint
is satisfied. We also satisfy the $\mbox{Br}[b\rightarrow s + \gamma]$
constraint and the Higgs mass constraint. The $S_{\phi K_S}$ values
shown in the table are the minimum that can be reached for given
$m_{1/2}$ and $A_0$.

It can be seen from Table I that the $S_{\phi K_S}$ values in mSUGRA
differ only slightly from the SM prediction which is $\sin 2 \beta$
evaluated using just the CKM phase. The branching ratios of $\bpk$
decays also do not differ much from the SM prediction. Even if one
went to the large weak annihilation region to accommodate the large
branching ratios, $S_{\phi K_S}$ would still be similar to the numbers
in Table I. Therefore, mSUGRA can not explain the large BR and the
2.7$\sigma$ difference between the $S_{\phi K_S}$ and the $S_{J/\Psi
  K_S}$ experimental results.
The reason is that, in mSUGRA, the only flavor violating source is in
the CKM matrix, which cannot provide enough flavor violation needed
for the $b\rightarrow s$ transition in $\bpk$ decays.  In the next
section, we will search for the minimal extension of mSUGRA that can
solve both the BR and CP problems of $\bpk$ decays.

\begin{table}[h]
 \begin{tabular}{|c|c|c|c|c|c|c|c|c|}
 \hline  $\tan\beta$ & \multicolumn{4}{|c|}{$10$} & \multicolumn{4}{|c|}{$40$} \\
 \hline $|A_0|$ & $800$ & $600$ & $400$& $0$ & $800$ & $600$ & $400$& $0$\\
 \hline  $m_{1/2}=400$ & $\ba{c} 0.74 \ea$ & $\ba{c} 0.74 \ea$ & $\ba{c} 0.73 \ea$ & $\ba{c} 0.73 \ea$ & $\ba{c} 0.71 \ea$ & $\ba{c} 0.70 \ea$ & $\ba{c} 0.70 \ea$ & $\ba{c} 0.69 \ea$ \\
 \hline  $m_{1/2}=500$ & $\ba{c} 0.74 \ea$ & $\ba{c} 0.74 \ea$ & $\ba{c} 0.74 \ea$ & $\ba{c} 0.74 \ea$ & $\ba{c} 0.72 \ea$ & $\ba{c} 0.72 \ea$ & $\ba{c} 0.72 \ea$ & $\ba{c} 0.71 \ea$\\
 \hline
 \end{tabular}
 \caption{$S_{\phi K_S}$ with small weak annihilation at $\tan\beta=10$ and 40 in mSUGRA. The values are the minimum that can be
 reached subject to all other experimental constraints\label{ms1}}
\end{table}

\section{SUGRA model with Non-universal A terms}
In the last section, we showed that mSUGRA contributions to $\bpk$
decays are negligible and thus mSUGRA needs to be extended if it is to
explain the experimental results of $\bpk$
decays. It is obvious that
some non-universal soft breaking terms which can contribute to the
$b\rightarrow s$ transition are necessary. There are two ways of
enhancing the mixing between the second and the third generation: one
can have non universal terms in the squark mass matrices or in the
$A^{U,D}$ matrices of Eq.(\ref{eq07}). However, in a GUT model, at
least the Standard Model gauge group must hold at $M_G$ and hence the
only squark $m^2_{23}$ that can occur is either left-left or
right-right coupling. As discussed in \cite{Kane03}, such
non-universal terms produce only small effects on $\bpk$ decays. Thus
we are led to models with left-right mixing which can occur in the
$A^{U,D}$ matrices as the simplest possible non-universal term
relevant to $\bpk$ decays. In this work then, we choose a model with
non-zero (2,3) elements in the trilinear coupling A terms of
Eq.({\ref{eq07}}) to enhance the left-right mixing of the second and
the third generation. The A terms with non-zero 23 elements can be
written as
\begin{equation}\label{eq14}
A^{U,D} = A_0 Y^{U,D} + \Delta A^{U,D}
\end{equation}
where $\Delta A^{U,D}$ are $3 \times 3$ complex matrices and $\Delta
A^{U,D}_{ij} = |\Delta A^{U,D}_{ij}| e^{\phi^{U,D}_{ij}}$. When
$\Delta A^{U,D} = 0$, mSUGRA is recovered.  For simplicity, we will
discuss first the case of non-zero $\Delta A^{D}_{23}$ and non-zero 
$\Delta A^{D}_{32}$ for $\tan\beta = 10$ and 40. In both cases, all
other entries in $\Delta A^{U,D}$ are set to zero. The other
parameters are same as in the mSUGRA case. We will set the phases such
that the EDM constraints are obeyed. At the GUT scale, we use a diagonal
Yukawa texture for $Y^U$, while $Y^D$ is constructed as $VY^D_d$ where
$V$ is the CKM matrix and $Y^D_d$ is the diagonalized matrix of the
down type Yukawa. The phenomenological requirements for the Yukawa
matrices are that they produce the correct quark masses and the
correct CKM matrix. Any other Yukawa texture which satisfies the same
requirements can be obtained through unitary rotations. Therefore, our
results can be recovered with other Yukawa textures if our A terms are
rotated along with the Yukawas.

In the calculations of decay amplitudes, we will use QCD parameters
for the small weak annihilation region (see the last section) where the
theory is reliable. In general it is possible that (see \cite{Du02a}
for the calculational details of weak annihilation) the new physics
can change the behavior of annihilation contributions when the
relevant Wilson coefficients can be reduced or increased
significantly. However, in our case with non-zero $\Delta
A^{U,D}_{23,32}$ terms, the SUSY contribution mainly affects the
Wilson coefficients $C_{8g(7\gamma)}$ (possibly also
$\tilde{C}_{8g(7\gamma)}$) and these coefficients will not change the
annihilation contributions compared to what we have in the SM
calculation and thus our previous conclusion about the annihilation
terms still holds.

\vspace{3ex}{\bfseries Case I: $|\Delta A^{D}_{23}|=|\Delta
  A^{D}_{32}|$ and $\phi ^{D}_{23}\neq\phi^{D}_{32}$}\vspace{3ex}

\begin{table}[h]
 \begin{tabular}{|c|c|c|c|c|c|}
 \hline $|A_0|$ & $800$ & $600$ & $400$& $0$ & $|\Delta A^{D}_{23(32)}|$\\
 \hline  $m_{1/2}=300$ & $\ba{c} -0.50 \ea$ & $\ba{c} -0.49 \ea$ & $\ba{c} -0.47 \ea$ & $\ba{c} -0.43 \ea$ & $\ba{c} \sim 50 \ea$\\
 \hline  $m_{1/2}=400$ & $\ba{c} -0.43 \ea$ & $\ba{c} -0.40 \ea$ & $\ba{c} -0.38 \ea$ & $\ba{c} -0.36 \ea$ & $\ba{c} \sim 110 \ea$\\
 \hline  $m_{1/2}=500$ & $\ba{c} -0.46 \ea$ & $\ba{c} -0.46 \ea$ & $\ba{c} -0.44 \ea$ & $\ba{c} -0.31 \ea$ & $\ba{c} \sim 200 \ea$\\
 \hline  $m_{1/2}=600$ & $\ba{c} -0.15 \ea$ & $\ba{c} -0.13 \ea$ & $\ba{c} -0.04 \ea$ & $\ba{c} 0.05 \ea$ & $\ba{c} \sim 280 \ea$\\
 \hline
 \end{tabular}
 \caption{$S_{\phi K_S}$ at $\tan\beta=10$ with non-zero $A^D_{23}$ and $A^D_{32}$.}\label{DA1}
\end{table}
\begin{table}[h]
 \begin{tabular}{|c|c|c|c|c|c|c|c|c|}
 \hline $|A_0|$ & \multicolumn{2}{|c|}{$800$} & \multicolumn{2}{|c|}{$600$} &
\multicolumn{2}{|c|}{$400$} &
\multicolumn{2}{|c|}{$0$}\\
 \hline  $m_{1/2}=300$ & $\ba{c} 1.2\% \ea$ & $\ba{c} -3.7\% \ea$ & $\ba{c}
1.4\% \ea$ & $\ba{c} -3.6\% \ea$ & $\ba{c}
1.7\% \ea$ & $\ba{c} -3.6\% \ea$ & $\ba{c} 2.2\% \ea$ & $\ba{c} -3.5\% \ea$\\
 \hline  $m_{1/2}=400$ & $\ba{c} 1.9\% \ea$ & $\ba{c} -3.5\% \ea$ & $\ba{c}
2.0\% \ea$ & $\ba{c} -3.4\% \ea$ & $\ba{c}
2.2\% \ea$ & $\ba{c} -3.3\% \ea$ & $\ba{c} 2.3\% \ea$ & $\ba{c} -3.3\% \ea$\\
 \hline  $m_{1/2}=500$ & $\ba{c} 2.6\% \ea$ & $\ba{c} -3.5\% \ea$ & $\ba{c}
2.6\% \ea$ & $\ba{c} -3.6\% \ea$ & $\ba{c}
2.5\% \ea$ & $\ba{c} -3.5\% \ea$ & $\ba{c} 2.4\% \ea$ & $\ba{c} -3.2\% \ea$\\
 \hline  $m_{1/2}=600$ & $\ba{c} 2.0\% \ea$ & $\ba{c} -2.8\% \ea$ & $\ba{c}
2.1\% \ea$ & $\ba{c} -2.7\% \ea$ & $\ba{c}
2.1\% \ea$ & $\ba{c} -2.5\% \ea$ & $\ba{c} 2.2\% \ea$ & $\ba{c} -2.2\% \ea$\\
 \hline
 \end{tabular}
 \caption{$\mathcal{A}
 _{b\rightarrow s + \gamma}$ (left) and $\mathcal{A}_{\phi K^{\mp}}$ (right) at
$\tan\beta=10$ with non-zero $A^D_{23}$ and $A^D_{32}$.\label{DA2}}
\end{table}

We show our results for $|\Delta A^{D}_{23}|=|\Delta A^{D}_{32}|$ but
$\phi ^{D}_{23}\neq\phi^{D}_{32}$ with $\tan\beta = 10$ in Table
\ref{DA1}.  We note that $|\Delta A^{D}_{23(32)}|$ is an increasing
function of $m_{1/2}$. The phases $\phi^{D}_{23}$ and $\phi^{D}_{32}$
are approximately $-30^{\circ}$ and $(75 \sim 115)^{\circ}$,
respectively. The other SUSY phases are: $\phi_1 \sim 22^{\circ}$,
$\phi_3 \sim 31^{\circ}$ and $\phi_\mu \sim -11^{\circ}$. In addition,
as mentioned above, the phase of $A_0$, i.e. $\alpha_A$, is set to be
$\pi$.

The $\mbox{Br}[\bpkm]$ is $\sim 10 \times 10^{-6}$ in the parameter
space of Table II. We satisfy all other experiment constraints. We
see that SUGRA models can explain the large BR and $S_{\phi K_S}$ of
the $\bpk$ decay modes even in the small annihilation region. Comparing
with Eq.(\ref{eq01}), one sees that the values of $S_{\phi K_S}$ in
the Table are within $1\sigma$ range of the experimental measurement.
Reducing the $\mbox{Br}[\bpkm]$ allows one to lower $S_{\phi K_s}$.
For example, for $A_0=0$ and $m_{1/2}=600$ GeV, by adjusting
$\phi^{D}_{32}$, $S_{\phi K_s}$ can be reduced to -0.05 with
$\mbox{Br}[\bpkm] \sim 9 \times 10^{-6}$.
In Table \ref{DA2} we show the direct CP asymmetries of the $\bpkm$ decay,
i.e.  $\mathcal{A}_{\phi K^{\mp}}$, using the same parameters as Table
\ref{DA1}.  The CP asymmetry is around $-(2\sim 3)\%$ and agrees with
the experimental result shown in Eq.(\ref{eq01b}).  This prediction
depends on the choice of $\phi_{A,H}$ in Eq.(\ref{eq13}). For example,
if we use $\phi_{A,H}=28^{\circ}$, we generate a large
$\mathcal{A}_{\phi K^{\mp}}\sim 27\%$. We find that there exists a
reasonably large range of $\phi_{A,H}$ where we can satisfy the
current bound on $\mathcal{A}_{\phi K^{\mp}}$. For example, at
$m_{1/2} = 300$ GeV and $A_0 = 800$ GeV where the SUSY contribution is the
largest, we find that $\mathcal{A}_{\phi K^{\mp}}$ varies from $9\%$
to $-4\%$ when $\phi_{A,H}$ varies from $-100^{\circ}$ to
$-50^{\circ}$ (for simplicity, we set $\phi_A = \phi_H$). In addition,
since  the annihilation contribution is small in that range, we find
that the branching ratio is around $(9.5\sim 11) \times 10^{-6}$. The
CP asymmetry of $b\rightarrow s \gamma$ is $\sim$1-3\%. The present
experimental errors for $C_{\phi K_S}$ are still large. For
this model, $C_{\phi K_S} \sim - \mathcal{A}_{\phi K^{\mp}}$,
which may be tested by future data.

\begin{table}[h]
 \begin{tabular}{|c|c|c|c|c|c|c|c|c|}
 \hline $|A_0|$ & \multicolumn{2}{|c|}{$800$} & \multicolumn{2}{|c|}{$600$} &
\multicolumn{2}{|c|}{$400$} &
\multicolumn{2}{|c|}{$0$}\\
 \hline  $m_{1/2}=300$ & $\ba{c} -0.40 \ea$ & $\ba{c} 10.0 \ea$ & $\ba{c} -0.38
\ea$ & $\ba{c} 10.0 \ea$ & $\ba{c} -0.33
\ea$ & $\ba{c} 10.1 \ea$ & $\ba{c} -0.05 \ea$ & $\ba{c} 10.0 \ea$\\
 \hline  $m_{1/2}=400$ & $\ba{c} -0.11 \ea$ & $\ba{c} 8.0 \ea$ & $\ba{c} -0.05
\ea$ & $\ba{c} 8.0 \ea$ & $\ba{c} 0.04
\ea$ & $\ba{c} 7.9 \ea$ & $\ba{c} 0.28 \ea$ & $\ba{c} 8.0 \ea$\\
 \hline  $m_{1/2}=500$ & $\ba{c} 0.07 \ea$ & $\ba{c} 6.0 \ea$ & $\ba{c} 0.16
\ea$ & $\ba{c} 6.1 \ea$ & $\ba{c} 0.24 \ea$
& $\ba{c} 6.1 \ea$ & $\ba{c} 0.37 \ea$ & $\ba{c} 6.2 \ea$\\
 \hline  $m_{1/2}=600$ & $\ba{c} 0.37 \ea$ & $\ba{c} 6.2 \ea$ & $\ba{c} 0.44
\ea$ & $\ba{c} 6.2 \ea$ & $\ba{c} 0.49 \ea$
& $\ba{c} 6.2 \ea$ & $\ba{c} 0.58 \ea$ & $\ba{c} 6.2 \ea$\\
 \hline
 \end{tabular}
 \caption{$S_{\phi K_S}$ (left) and $\mbox{Br}[\bpkm] \times 10^6$ (right) at
$\tan\beta=40$ with non-zero $\Delta
A^D_{23}$ and $\Delta A^D_{32}$.\label{DA3}}
\end{table}
\begin{table}[h]
 \begin{tabular}{|c|c|c|c|c|c|c|c|c|}
 \hline $|A_0|$ & \multicolumn{2}{|c|}{$800$} & \multicolumn{2}{|c|}{$600$} &
\multicolumn{2}{|c|}{$400$} &
\multicolumn{2}{|c|}{$0$}\\
 \hline  $m_{1/2}=300$ & $\ba{c} -6.3\% \ea$ & $\ba{c} -3.5\% \ea$ & $\ba{c}
-5.6\% \ea$ & $\ba{c} -3.4\% \ea$ & $\ba{c}
-5.2\% \ea$ & $\ba{c} -3.3\% \ea$ & $\ba{c} -3.6\% \ea$ & $\ba{c} -2.6\% \ea$\\
 \hline  $m_{1/2}=400$ & $\ba{c} -3.0\% \ea$ & $\ba{c} -3.0\% \ea$ & $\ba{c}
-2.1\% \ea$ & $\ba{c} -2.9\% \ea$ & $\ba{c}
-1.7\% \ea$ & $\ba{c} -2.6\% \ea$ & $\ba{c} -0.7\% \ea$ & $\ba{c} -1.8\% \ea$\\
 \hline  $m_{1/2}=500$ & $\ba{c} -0.5\% \ea$ & $\ba{c} -2.9\% \ea$ & $\ba{c}
-0.4\% \ea$ & $\ba{c} -2.5\% \ea$ & $\ba{c}
-0.2\% \ea$ & $\ba{c} -2.2\% \ea$ & $\ba{c} 0.2\% \ea$ & $\ba{c} -1.7\% \ea$\\
 \hline  $m_{1/2}=600$ & $\ba{c} 0.2\% \ea$ & $\ba{c} -1.7\% \ea$ & $\ba{c}
0.3\% \ea$ & $\ba{c} -1.4\% \ea$ & $\ba{c}
0.4\% \ea$ & $\ba{c} -1.2\% \ea$ & $\ba{c} 0.6\% \ea$ & $\ba{c} -0.8\% \ea$\\
 \hline
 \end{tabular}
 \caption{$\mathcal{A}
 _{b\rightarrow s + \gamma}$ (left) and $\mathcal{A}_{\phi K^{\mp}}$ (right) at
$\tan\beta=40$ with non-zero $A^D_{23}$ and $A^D_{32}$.\label{DA4}}
\end{table}

In Table \ref{DA3} and Table \ref{DA4}, we give our results for $\tan\beta = 40$ with
non-zero $\Delta A^D_{23(32)}$. The phases $\phi^{D}_{23}$ and
$\phi^{D}_{32}$ are $-(70\sim 0)^{\circ}$ and $(80\sim 110)^{\circ}$,
respectively. $\phi_1 \sim (25 \sim 60)^{\circ}$, $\phi_3 \sim
25^{\circ}$ and $\phi_\mu \sim -8^{\circ}$.  The off-diagonal elements
$|\Delta A^D_{23(32)}|$ vary from 90 GeV to 250 GeV as $m_{1/2}$ increases.
We compare Table \ref{DA3} with the results for $\tan\beta=10$ shown in Table
\ref{DA1} and we see that only low $m_{1/2}$ can satisfy experimental
data for $\tan\beta=40$. The most important reason for this is that
left-right mixing of the second and the third generation decreases
significantly with increasing $\tan\beta$. This comes about as
follows. The RGE running of $A^D_{23(32)}$ is not sensitive to
$\tan\beta$. Therefore, for the same size of $A^D_{23(32)}$ input at
the GUT scale, the weak scale values of $A^D_{23(32)}$ do not differ
much for different $\tan\beta$. However, the $A^D$ term enters into the
down squark matrix after electroweak symmetry breaking when $H_1$ (see
Eq.(\ref{eq07})) grows a vacuum expectation value proportional to
$\cos\beta$. Hence the left-right mixing
between the second and the third generation in the down squark matrix
will be smaller for large $\tan\beta$. For low $m_{1/2}$ this reduction
can be compensated by increasing the magnitude of $A^{D}_{23(32)}$.
For example, for  $m_{1/2} = 300$ GeV, we use $|A^{D}_{23(32)}| \sim 90$ GeV
in this case compared to 50 GeV in the case of $\tan\beta = 10$ (see
Table \ref{DA1}). The chargino diagram contribution increases with
$\tan\beta$ and can help to generate a large BR. But for large
$m_{1/2}$, when the chargino contribution goes down, 
$|A^{D}_{23(32)}|$ must become much larger. However, as $A_{23(32)}$
increases, the pseudoscalar Higgs mass becomes small at the same time
(but $\mu$ does not get smaller), which prevents $|A^{D}_{23(32)}|$
from having an unlimited increase. For example, for $m_{1/2} = 600$ GeV
and $A_0 = 800$ GeV, $|A^{D}_{23(32)}| = 250$ GeV generates $m_A = 580$
GeV which is still allowed for the dark matter constraint to be
satisfied in the $\tilde{\tau}\leftrightarrow\tilde{\chi}^0$
co-annihilation channel.  If $|A^{D}_{23(32)}|$ is increased more, the
pseudoscalar mass gets smaller and the dark matter constraint can
still be satisfied due to the available $\chi^0_1\chi^0_1\rightarrow
A\rightarrow f\bar f$ channel. But with a further reduction of the
pseudoscalar mass by increasing $|A^{D}_{23(32)}|$ further, this
channel goes away when $m_A < 2 m_{\tilde{\chi}^0}$ and we must again
satisfy the relic density using the stau-neutralino co-annihilation
channel. However, the improvement of $S_{\phi K_s}$ in this scenario
is small. For example, for the point mentioned above,
$|A^{D}_{23(32)}|$ can be increased to around 480 GeV with relic
density in the $\tilde{\tau}\leftrightarrow\tilde{\chi}^0$ channel but
$S_{\phi K_S}$ can only be reduced from 0.37 (see Table \ref{DA3}) to
0.22 with the same branching ratio. Thus, the $S_{\phi K_S}$ and the
branching ratio still cannot be satisfied.

If we use $\phi^{D}_{23} = \phi^{D}_{32}$ (equal phases) we have one
less parameter, but that choice will not be able to satisfy
experimental results. The reason is that the weak phase from the gluino
contributions in the Wilson coefficients $C_{8g}$ and the weak phase from $\tilde{C}_{8g}$
will cancel when $\phi^{D}_{23} = \phi^{D}_{32}$ because
$C_{8g}$ depends on $A^D_{23}$ but $\tilde{C}_{8g}$ depends on
$(A^D_{32})^\star$. For example, for $\tan\beta = 10$ we find that
$S_{\phi K_S} \sim 0.7$ since the gluino contribution dominates at
lower $\tan\beta$. At $\tan\beta = 40$, $S_{\phi K_S}$ can reach 0.45
since the chargino contribution is larger at higher $\tan\beta$, but
this is not enough to satisfy the data.

\vspace{3ex}{\bfseries Case II: $|\Delta A^{U}_{23}|=|\Delta
  A^{U}_{32}|$ and $\phi ^{U}_{23} = \phi^{U}_{32}$}\vspace{3ex}

\begin{table}[h]
 \begin{tabular}{|c|c|c|c|c|c|c|c|c|c|}
 \hline $|A_0|$ & \multicolumn{2}{|c|}{$800$} & \multicolumn{2}{|c|}{$600$} &
\multicolumn{2}{|c|}{$400$} &
\multicolumn{2}{|c|}{$0$} & $|\Delta A^{U}_{23(32)}| $ (GeV)\\
 \hline  $m_{1/2}=300$ & $\ba{c} 0.03 \ea$ & $\ba{c} 8.4 \ea$ & $\ba{c} 0.04
\ea$ & $\ba{c} 9.0 \ea$ & $\ba{c} 0.01 \ea$ & $\ba{c} 8.0 \ea$ & $\ba{c} 0.17 \ea$ & $\ba{c} 8.0 \ea$ & $\ba{c} \sim 300 \ea$ \\
 \hline  $m_{1/2}=400$ & $\ba{c} -0.07 \ea$ & $\ba{c} 8.5 \ea$ & $\ba{c} -0.03
\ea$ & $\ba{c} 8.4 \ea$ & $\ba{c} 0
\ea$ & $\ba{c} 7.1 \ea$ & $\ba{c} 0.32 \ea$ & $\ba{c} 6.3 \ea$ & $\ba{c} \sim
600 \ea$ \\
 \hline  $m_{1/2}=500$ & $\ba{c} 0 \ea$ & $\ba{c} 6.5 \ea$ & $\ba{c} 0.07
\ea$ & $\ba{c} 6.4 \ea$ & $\ba{c} 0.18 \ea$ & $\ba{c} 6.0 \ea$ & $\ba{c} 0.44 \ea$ & $\ba{c} 6.1 \ea$ & $\ba{c} \sim 800 \ea$ \\
 \hline  $m_{1/2}=600$ & $\ba{c} 0.27 \ea$ & $\ba{c} 6.1 \ea$ & $\ba{c} 0.30
\ea$ & $\ba{c} 6.1 \ea$ & $\ba{c} 0.35 \ea$ & $\ba{c} 6.1 \ea$ & $\ba{c} 0.51 \ea$ & $\ba{c} 5.9 \ea$ & $\ba{c} \sim 1000 \ea$ \\
 \hline
 \end{tabular}
 \caption{$S_{\phi K_S}$ (left) and $\mbox{Br}[\bpkm] \times 10^6$ (right) at
$\tan\beta=40$ with non-zero $\Delta
A^U_{23}$ and $\Delta A^U_{32}$.\label{UA1}}
\end{table}

\begin{table}[h]
 \begin{tabular}{|c|c|c|c|c|c|c|c|c|c|}
 \hline $|A_0|$ & \multicolumn{2}{|c|}{$800$} & \multicolumn{2}{|c|}{$600$} &
\multicolumn{2}{|c|}{$400$} &
\multicolumn{2}{|c|}{$0$}& $|\Delta A^{U}_{23(32)}| $ (GeV)\\
 \hline  $m_{1/2}=300$ & $\ba{c} 0.17 \ea$ & $\ba{c} 6.5 \ea$ & $\ba{c} 0.16
\ea$ & $\ba{c} 6.3 \ea$ & $\ba{c} 0.32 \ea$
& $\ba{c} 6.1 \ea$ & $\ba{c} 0.60 \ea$ & $\ba{c} 5.2 \ea$ & $\ba{c} \sim 300
\ea$ \\
 \hline  $m_{1/2}=400$ & $\ba{c} 0.37 \ea$ & $\ba{c} 4.7 \ea$ & $\ba{c} 0.39
\ea$ & $\ba{c} 4.6 \ea$ & $\ba{c} 0.46 \ea$
& $\ba{c} 4.3 \ea$ & $\ba{c} 0.62 \ea$ & $\ba{c} 4.3 \ea$ & $\ba{c} \sim 550
\ea$ \\
 \hline
 \end{tabular}
 \caption{$S_{\phi K_S}$ (left) and $\mbox{Br}[\bpkm] \times 10^6$ (right) at $\tan\beta=10$ with non-zero $\Delta A^U_{23}$ and $\Delta A^U_{32}$.\label{UA2}}
\end{table}

In this section we discuss the case $\Delta A^D_{23(32)} = 0$ but
$\Delta A^U_{23(32)} \neq 0$. The phases used are similar to those
used in first two cases except $\phi ^{U}_{23} = \phi^{U}_{32}$. This case is more complicated than the
$A^D_{23(32)} \neq 0$ case. We find that it is easier to start by comparing
them.

The first important change is that the $\Delta A^U_{32}$ contribution
is much smaller than the $\Delta A^D_{32}$ contribution to the mixing
between the second and the third generation in the down squark mass
matrix due to the suppression by the second generation Yukawa
coupling in the RGE of $A^D_{32}$. (Thus our choice of $\phi ^{U}_{23} = \phi^{U}_{32}$ has no loss of generality.) Consequently, the size of the
Wilson coefficient $\tilde{C}_{8g}$ is significantly reduced. Although
$\Delta A^U_{23}$ still contributes, that contribution is also reduced (compared to $\Delta A^D_{23}$)
due to the RGE. Therefore, compared with the first case the total SUSY contributions 
are reduced especially for
$\tan\beta = 10$ and thus it becomes harder to fit the experimental results.

Another important change is the roles of some experimental constraints
which are not important in the first case in the sense that they do
not prevent the SUSY contributions from increasing, or at least their
limits are not reached when we have solutions satisfying the B-decay
data. Below are some comments concerning this:

\noindent 1. For $\tan\beta = 40$ and low $m_{1/2}$, i.e. 300 GeV, the
$\mbox{Br}[\bsg]$ will constrain the size of $\Delta A^U_{23(32)}$ .
This is why the $S_{\phi K_S}$ and the branching ratio fit is not as
good as the corresponding one shown in Table \ref{DA1} for the
$A^D_{23(32)} \neq 0$ case.

\noindent 2. When $m_{1/2}$ increases, the size of $\Delta A^U_{23(32)}$ also
needs to be increased. But three other additional constraints are
present, i.e $\Delta M_K$ and $\epsilon_K$ from the $K^0-\overline{K^0}$ 
mixing and the mass of smallest up squarks (right-handed stop)
$m_{\tilde{t}_R}$. 
For example, for $m_{1/2} = 500$ and $A_0 = 600$ (and $m_0 = 431$ GeV
by the relic density constraint) we get $m_{\tilde{g}} \sim
m_{\tilde{q}} \sim 1000$ GeV (where $m_{\tilde{q}}$ is the average
squark mass and $m_{\tilde{g}}$ is the mass of the gluino, see
\cite{Ciu} for more details) and we find that
$\sqrt{|\mbox{Re}(\delta^d_{12})^2_{LL}|} = 7.1 \times 10^{-2}$ and
$\sqrt{|\mbox{Im}(\delta^d_{12})^2_{LL}|} = 9.7 \times 10^{-3}$ which
are allowed by the experimental bounds on $\Delta M_K$ and $\epsilon_K$ \cite{Ciu} (the sizes of
$(\delta^d_{12})_{LR}$, $(\delta^d_{12})_{RL}$ and
$(\delta^d_{12})_{RR}$ are around $10^{-8} \sim 10^{-7}$ and thus
these bounds can be safely ignored in our case).

\noindent 3. The situation for the right-handed stop mass is similar to the
pseudoscalar Higgs case we mentioned at the end of Case I.  We use the
$\tilde{\tau}\leftrightarrow\tilde{\chi}^0$ channel to satisfy the
dark matter constraints. Although it's possible to use a larger
$A^U_{23(32)}$ which consequently reduces $m_{\tilde{t}_R}$ more and
then opens the $\tilde{t}_R\leftrightarrow\tilde{\chi}^0$ channel, the
room is small due to the smallness of $m_{\tilde{\chi}^0}$. In
addition, the $M_K$ and the $\epsilon_K$ bounds become harder to
satisfy when $m_{\tilde{t}_R}$ is small. Therefore, as in the case where the  pseudoscalar Higgs mass becomes small, possible improvements can not satisfy
the experimental results of both $S_{\phi K_S}$ and
$\mbox{Br}[\bpkm]$.

A further difference is the $\tan\beta$ dependence. In Case I, as was
discussed above, gluino contributions depend inversely on $\tan\beta$
due to the way that $A^D_{23(32)}$ enters into the down squark mass
matrix.  But in this case, the gluino contributions are reduced
significantly and the chargino plays a more important role, which will
be enhanced by $\tan\beta$. Therefore, in this case, we see that
larger $\tan\beta$ can satisfy the experimental results at small
$m_{1/2}$, but small $\tan\beta$ cannot and that is why we have only
given results in Table \ref{UA2} for two values of $m_{1/2}$ at
$\tan\beta = 10$ since higher $m_{1/2}$ cannot improve the situation.

We also comment concerning $A_0$ and its phase. So far,
we have used the phase $\pi$ for $A_0$. We find that using a different
phase will not improve the results greatly. In general, the
improvement is at a few percent level. (This holds also for case I.) For example, in Case II, for
$\tan\beta = 40$, $m_{1/2} = 400$ and $A_0 = 800$, we find that using
$\alpha_A \sim -95^{\circ}$ can improve $S_{\phi K_S}$ from -0.04 to
-0.06.

Finally we note that the values of $\mathcal{A}
 _{b\rightarrow s + \gamma}$ and $\mathcal{A}_{\phi K^{\mp}}$ remain small, i.e.  $\mathcal{A}
 _{b\rightarrow s + \gamma}$ and $\mathcal{A}_{\phi K^{\mp}}$ are $-(3 \sim 0) \%$ and $-(3\sim 1) \%$ at $\tan\beta=10$, and $-(5 \sim 0) \%$ and $-(3\sim 1) \%$ at  $\tan\beta=40$.


\section{Conclusion}
In this paper, we have probed the $\bpk$ decays in SUGRA models with
CP violating phases to explain the discrepancy between the
experimental measurements and the SM predictions of the CP asymmetry of
$\bpks$ and the branching ratios of the $\bpk$ decays.  We have calculated
the CP asymmetries of $\bpkm$ and $\bsg$. In our analysis, we
implemented all relevant experimental constraints, e.g.
$\mbox{Br}[\bsg]$, relic density, $K^0 - \bar{K^0}$ mixing parameters
and electron and neutron EDMs.  We used the improved QCD factorization
method \cite{bbns, Du02a} for the calculation of decay amplitudes.

The SM not only can not explain the CP asymmetry of $\bpks$, it also
fails to satisfy the $\mbox{Br}[\bpk]$ data barring the region of
large weak annihilation where the theory is most suspect. We then
studied the mSUGRA model and found that it also has the same problem.
Therefore, if the current experimental results continue to hold in the
future, it will signal the first significant breakdown of the Standard
Model and also of mSUGRA. This conclusion is important in the sense
that one needs to construct a more complicated SUGRA model to satisfy
experimental measurements which will provide important guidance to our
future research on SUSY models and their signals at the accelerator
experiments.

We have considered the extension of the mSUGRA model by adding
non-universal A terms. For a GUT theory, the only natural choice is to
have a left-right mixing between the second and the third generation
in the up or down quark sectors i.e. $\Delta A^{U,D}_{23}$ and $\Delta
A^{U,D}_{32}$ terms. We have examined thoroughly several different
possibilities in this extension and their theoretical predictions and
have found a large region of parameter space where all experimental
results can be satisfied, including the CP asymmetries and branching
ratios of the $\bpk$ decays.  This result is obtained without
resorting to large weak annihilation amplitudes and so is based on
reliable calculations of hadronic decays, and thus provides useful hints
for the study of hadronic B decays. Further the size of $\Delta
A^{U,D}_{23}$ needed is the same as the other soft breaking terms, and
so is not anomalously small or large. Thus, this study also can
provide important phenomenological information not only for
accelerator physics but also for building models at the GUT scale and
for exploring physics beyond it. In this connection, models based on
Horava-Witten M-theory can naturally give rise to non-zero values of
$\Delta A_{23}$. In previous work \cite{ADHW} it was shown that it was
possible to construct a three generation model with $SU(5)$ symmetry
using a non-standard embedding based on a torus fibered Calabi-Yau
three fold with a del Pezzo base $dP_7$. The model allowed Wilson line
breaking to the Standard Model at $M_G$, and also had vanishing
instanton charges on the physical orbifold plane. If in addition one
assumed that the 5-branes in the bulk clustered around the distant
plane, one could explain without undue fine tuning the general
structure of the quark and lepton mass hierarchies and obtain the LMA
solution for neutrino oscillations \cite{ADHW,ADH,ADH1}. One can show
that the model naturally gives rise to $\Delta A_{23}$ non-universal
terms as required by the current B-factory data, while keeping the
squark mass matrices essentially universal. This model will be
discussed elsewhere \cite{ADH2}.

\begin{acknowledgments}
We thank Guohuai Zhu and Sechul Oh for a discussion on QCD factorization. 
This work is supported in part by a National Science Foundation Grant
PHY-0101015 and in part by  Natural Sciences and Engineering Research Council 
of Canada. 
\end{acknowledgments}

\end{document}